%% file: paper.tex
\documentclass[runningheads]{llncs}
\usepackage{lmodern}  
\usepackage[T1]{fontenc}  
\usepackage{graphicx}
\usepackage{xcolor}
\usepackage{url}
\usepackage[acronym,nonumberlist,nopostdot,notree,nomain]{glossaries}
\usepackage[nocompress]{cite}
\usepackage{orcidlink}

\newcommand{\chairmarker}[1]{%
  \begingroup\normalfont
  \includegraphics[height=1.4\fontcharht\font`\W]{src/img/markers/#1}%
  \endgroup
}
\input{acronyms}

\begin{document}
\title{Recent Advances in Data-Driven Business Process Management}
\titlerunning{Recent Advances in Data-Driven BPM}
%
\author{Ackermann, Lars\inst{1}\orcidlink{0000-0002-6785-8998} \and
Käppel, Martin\inst{1}\orcidlink{0009-0003-3420-8037} \and
Marcus, Laura\inst{8}\orcidlink{0000-0001-8786-081X} \and
Moder, Linda\inst{2}\orcidlink{0000-0003-2362-4839} \and
Dunzer, Sebastian\inst{3}\orcidlink{0000-0001-5229-1864} \and
Hornsteiner, Markus\inst{5}\orcidlink{0000-0002-8024-1220} \and
Liessmann, Annina\inst{3} \and
Zisgen, Yorck\inst{7}\orcidlink{0000-0002-9646-2829} \and
Empl, Philip\inst{5} \and
Herm, Lukas-Valentin\inst{4} \and
Neis, Nicolas\inst{4} \and
Neuberger, Julian\inst{1} \and
Poss, Leo\inst{5} \and
Schaschek, Myriam\inst{4} \and
Weinzierl, Sven\inst{3} \and
Wördehoff, Niklas\inst{8} \and
Jablonski, Stefan\inst{1} \and
Koschmider, Agnes\inst{7}\orcidlink{0000-0001-8206-7636} \and
Kratsch, Wolfgang\inst{8}\orcidlink{0000-0001-9815-0653} \and
Matzner, Martin\inst{3}\orcidlink{0000-0001-5244-3928} \and
Rinderle-Ma, Stefanie\inst{6}\orcidlink{0000-0001-5656-6108} \and
Röglinger, Maximilian\inst{2}\orcidlink{0000-0003-4743-4511} \and
Schönig, Stefan\inst{5}\orcidlink{0000-0002-7666-4482} \and
Winkelmann, Axel\inst{4}}
\authorrunning{Ackermann et al.}
%
\institute{Chair for Databases and Information Systems, University of Bayreuth, Germany 
\email{\{lars.ackermann,martin.kaeppel,julian.neuberger,stefan.jablonski\}\\@uni-bayreuth.de} 
\and
FIM Research Center, University of Bayreuth; Branch Business \& Information Systems Engineering of the Fraunhofer FIT, Bayreuth, Germany\\
\email{\{linda.moder, maximilian.roeglinger\}@fim-rc.de} \\
\and
Chair of Digital Industrial Service Systems, Friedrich-Alexander
University Erlangen-Nürnberg, Germany\\
\email{\{sebastian.dunzer,annina.liessmann,sven.weinzierl,martin.matzner\}@fau.de}
\and
Chair of Management and Information Systems, Julius-
Maximilians-University Würzburg, Germany\\
\email{\{myriam.schaschek,nicolas.neis,axel.winkelmann\}@uni-wuerzburg.de}
\and
Department for Information Systems, University of Regensburg, Germany\\
\email{\{markus.hornsteiner,philip.empl,leo.poss,stefan.schoenig\}@ur.de}
\and
Technical University of Munich, Germany; TUM School of
Computation, Information and Technology\\
\email{stefanie.rinderle-ma@tum.de}
\and
Group Business Informatics and Process Analytics, University of Bayreuth, Germany\\
\email{\{yorck.zisgen,agnes.koschmider\}@uni-bayreuth.de}
\and
FIM Research Center, University of Applied Sciences Augsburg; Branch Business \& Information Systems Engineering of the Fraunhofer FIT, Augsburg, Germany\\
\email{\{laura.marcus,niklas.woerdehoff,wolfgang.kratsch\}@fim-rc.de}
}
\maketitle              
\begin{abstract}
The rapid development of cutting-edge technologies, the increasing volume of data and also the availability and processability of new types of data sources has led to a paradigm shift in data-based management and decision-making. Since business processes are at the core of organizational work, these developments heavily impact \gls{bpm} as a crucial success factor for organizations. In view of this emerging potential, data-driven business process management has become a relevant and vibrant research area. Given the complexity and interdisciplinarity of the research field, this position paper therefore presents research 
insights regarding data-driven \gls{bpm}. 
\end{abstract}
%
%
\section{Introduction}
The rapid progress of cutting-edge digital technologies has ushered in a paradigm shift for data-based management and decision-making. On the one hand, more and more data in higher volume, variety, and velocity has become available ~\cite{Cappa}. This entails structured as well as unstructured data such as video or sensor data ~\cite{KRATSCH2022}, shedding light on a vast range of activities and context information. On the other hand, progress in the field of \gls{ai}, particularly \gls{genai}, is accelerating at an increasingly fast pace, as can be observed from the current hype around conversational \glspl{llm} ~\cite{Feuerriegel.2023}.

This creates a wide range of opportunities to inform and support data-driven management. Inspired by Christensen's disruptive innovation theory on the transformative impact of technologies on established industries and business models ~\cite{christensen2013disruptive}, these advancements have the potential to reshape industries and provide novel avenues for both organizational and societal progress. In this context, new data sources and efficient algorithms promise opportunities for the execution and management of business processes.

Given that business processes are at the core of organizational work~\cite{dumas2018fundamentals,Oberdorf}, \gls{bpm} is an essential success factor for organizations~\cite{dumas2018fundamentals}. \gls{bpm} thereby strongly benefits from recent technological advancements. Taking one step further from established approaches in process discovery and analysis in the context of process mining, data-driven \gls{bpm} can leverage both available data and tools utilizing their inherent potential to shift the current automation frontier further, while simultaneously increasing transparency. Through the increasing availability of data, blind spots can be uncovered, enabling organizations to better monitor, control, and adapt their business processes through extensive insights~\cite{KRATSCH2022}. 
Better processing capabilities support, complement, and replace human capacities and skills. They enable data-driven decisions, leading to more accurate and less subjective assessments, automated knowledge retention, faster decision-making, and implementation with fewer resources~\cite{Beverungen.2021,Kerpedzhiev.2021}. 

Even though there is vast potential, the research field of data-driven \gls{bpm}, i.e., a paradigm shift towards data-based decision-making and actions in \gls{bpm} driven by the progress of technology and data availability, is still in its infancy. Thus, it is of utmost importance for research to further explore the technological potential as well as its implications. 
In view of the complexity and contemporaneity of this research field, it is essential to approach data-driven \gls{bpm} from different perspectives. Various topics need to be considered, from the basics of data availability to the utilization perspective. Data-driven \gls{bpm} as a research field is so broad that it requires a collaboration including several disciplines to cover the entire research spectrum from technological foundations to socio-technical matters. Further, a strong link between research and practice is crucial. In our \gls{dproma}, we collaborate among eight research groups from Bavarian universities (for more details see Appendix A), conducting several workshops to discuss the state of the art and advances of data-driven \gls{bpm}. Based on the collective expert knowledge and advocating for a collaborative approach that brings together interdisciplinary expertise and networks, we developed this position paper to strengthen the knowledge on the topic of data-driven \gls{bpm} and provide insights into future research directions in this field. In this position paper, we therefore address the question: \emph{How can we take data-driven \gls{bpm} further?} Thereby, we focus on five distinct research fields within data-driven \gls{bpm}. 

The remainder of this paper is structured as follows. In Section \ref{sec:data-driven_bpm} we explain the overarching research framework for data-driven \gls{bpm}. Next, we summarize five research areas within data-driven \gls{bpm}, with a focus on the motivation for each field, the current state of the art, and directions for future research. We conclude with an Outlook and Call for Action in Section \ref{sec:outlook}. 

\section{Research Challenges in Data-Driven \gls{bpm}}
\label{sec:data-driven_bpm}
This section presents research challenges of data-driven \gls{bpm}. For this purpose, first the \gls{bpm} lifecycle is introduced, which lays the foundation to structurally formulate research directions. Second, six research topics are associated with the lifecycle steps they contribute to. These topics are then systematically grouped into five distinct research fields and presented in detail. Throughout this, we explore their practical relevance, current advancements, and challenges for future research. 
 
\subsection{Navigating the Process Lifecycle in Data-driven Process Management}
\begin{figure}[t]
    \centering
    \includegraphics[scale=0.60]{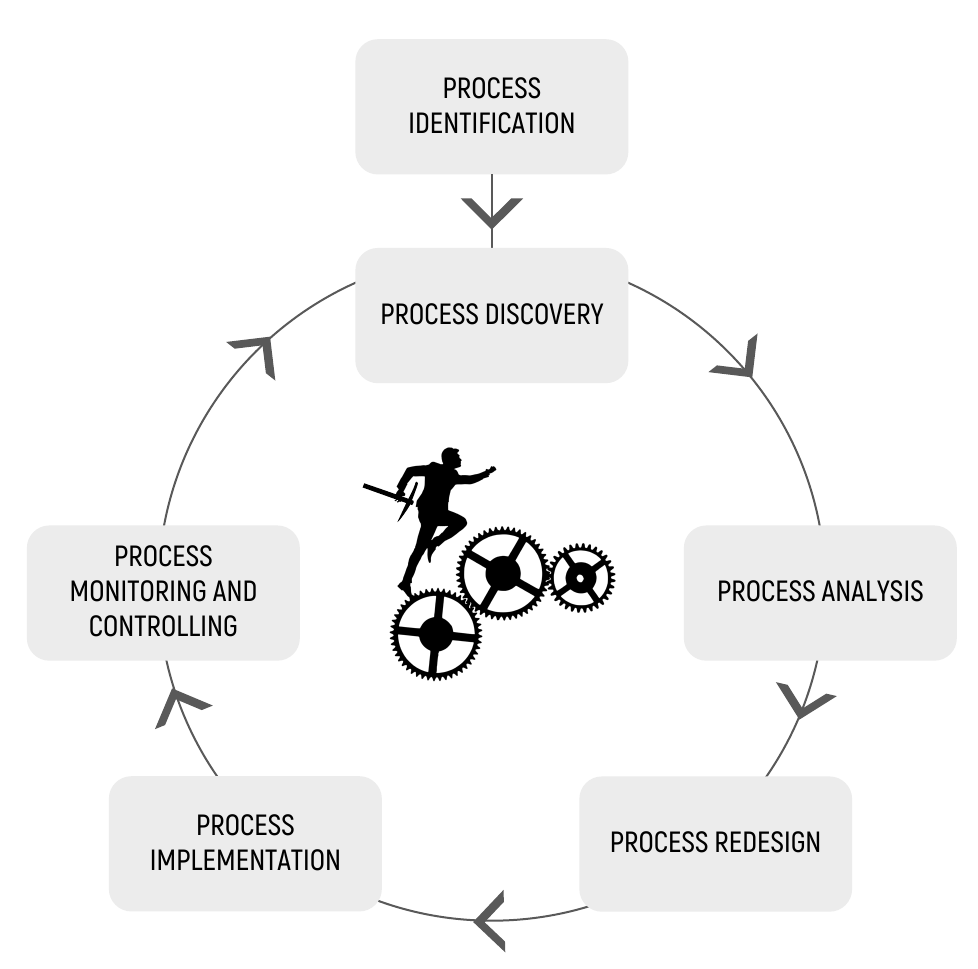}
    \caption{Process lifecycle according to Dumas et al.~\cite{dumas2018fundamentals}}
    \label{fig:process-lifecycle}
\end{figure}
The steps of \gls{bpm} commonly consist of six phases (so-called \emph{process lifecycle}) (see Fig. \ref{fig:process-lifecycle})~\cite{dumas2018fundamentals}. 
The lifecycle starts with the process \emph{identification phase}. In this phase, the processes of the organization are identified, the interest groups are analyzed and the problem statement is formulated~\cite{dumas2018fundamentals}. The result is a list of an organization's processes and their relationships, which is used to select the processes to be managed in the further course of the lifecycle. The remaining phases are run through separately for each process. 

In the \emph{process discovery phase} a process model is discovered representing the current implementation of the managed process. Based on this model, the \emph{process analysis phase} identifies, documents, and prioritizes issues associated with the process. The insights gained regarding the strengths and weaknesses of the process are then used in the \emph{process redesign phase}~\cite{dumas2018fundamentals}. This phase aims to to identify changes that reduce the previously identified weaknesses and exploit opportunities. The intended changes are recorded in a so-called to-be process model. The subsequent \emph{process implementation phase} finally realizes this to-be process model and transforms it into an executable process model~\cite{dumas2018fundamentals}. During the execution of the process, its performance is continuously monitored (within the so-called \emph{process monitoring and controlling phase}). Any inefficiencies, errors, or deviations from the intended behavior are identified and attempts are made to rectify them by means of suitable interventions~\cite{dumas2018fundamentals}. 

\begin{figure}[t]
    \centering
    \includegraphics[width=\textwidth]{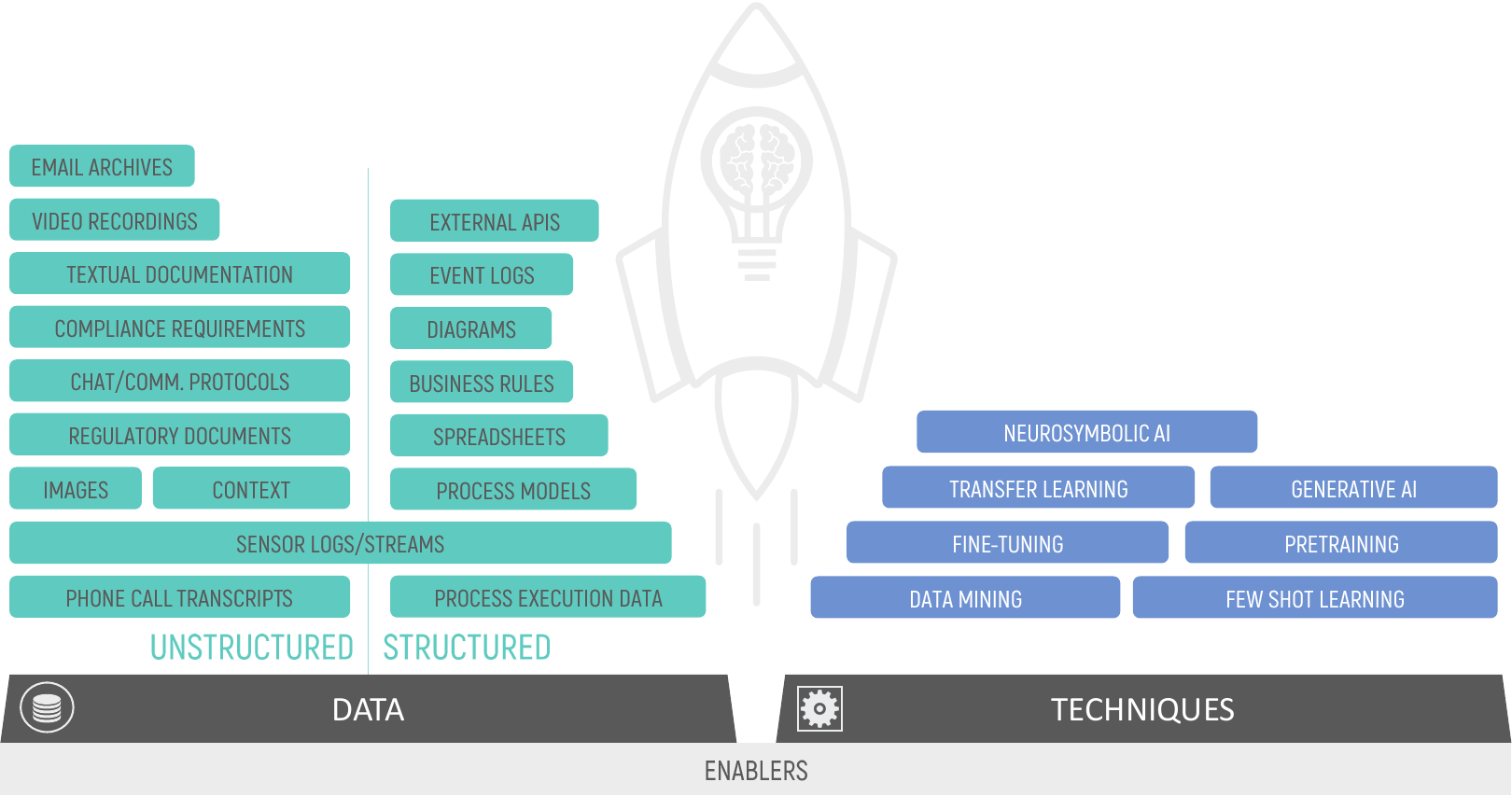}
    \caption{Enablers for innovation in data-driven \gls{bpm}}
    \label{fig:enablers}
\end{figure}

Since the lifecycle describes both the essential \gls{bpm} tasks and their interrelationships and due to its importance for \gls{bpm}, we use it as a natively structured basis for classifying current approaches to data-driven \gls{bpm}. Furthermore, through this classification (Section \ref{sec:data-processing} et seqq.), we illustrate the significance of these data-driven approaches for the lifecycle.

Progress in data-driven \gls{bpm} is made possible by newly available data sources and data types on the one hand and by advances in algorithms, particularly \gls{ai}, on the other~\cite{KRATSCH2022, Beverungen.2021,Kerpedzhiev.2021,rialti2018ambidextrous, czvetko2022data}. This includes - but is not limited to - the renaissance of \gls{dl} techniques and the emergence of \glspl{llm}. While process mining was almost exclusively limited to event logs for many years, numerous other data formats are now being processed. In addition to the increased availability of data, companies have also become more aware of the importance and value of data. Fig.~\ref{fig:enablers} gives a brief overview of the different forms of data that can now be processed in the lifecycle due to the emergence of suitable methods and techniques for automatic analysis and processing. In the following, we refer to both new types of data and new approaches to processing them as \emph{enablers}.

Process execution data is often available in a structured form but not necessarily in standard formats (e.g., as process event logs)~\cite{van2021event,wakup2015,vanEck2016sensorData,senderovich2016sensorData,koschmider2019,Mangler2023,Bellan2022,Ackermann2021,DBLP:conf/bpm/LepsienKK23}. Examples are system interaction logs covering records of low-level activities (e.g., single mouse clicks)~\cite{koschmider2019,van2021event,wakup2015}. Sensor data is a likewise low-level source of information and can be considered structured on the one hand (in terms of clear semantics of measurements) and unstructured on the other hand (in terms of information not directly accessible by algorithms, e.g., domain-dependent salient patterns)~\cite{vanEck2016sensorData,senderovich2016sensorData,Mangler2023}. In contrast to traditional process mining recent research activities in process mining not only incorporated structured data but also unstructured data such as natural language text (e.g., email archives, chat protocols or textual documentation)~\cite{Bellan2022,Ackermann2021} and multi-channel data (e.g., video recordings, images)~\cite{DBLP:conf/bpm/LepsienKK23,KRATSCH2022}. 

On the technical side of the enablers, it can be summarized that the latest innovations in deep learning, as in other areas, are a driving factor and a promising basis~\cite{tax2017predictive,harl2020explainable,metzger2014comparing,Kaeppel2023,Neuberger2023}. First and foremost, models from the field of generative AI are considered to be the best way to process complex data. However, unlike in image processing, for example, there is an acute lack of data~\cite{Kaeppel2021a,Kaeppel2021b, Kaeppel2023,Neuberger2023}, which is due to the increased data requirements of deep learning models. For these reasons, the focus is also on techniques such as transfer learning~\cite{Zhao2023}, pre-training~\cite{devlin2018bert}, and fine-tuning, the common goal of which is to reduce the data requirements of \gls{dl} models. However, due to challenges arising with aforementioned novel data sources established data mining techniques are used, for instance, to bridge the gap between low-level data and information relevant to \gls{bpm} (e.g., clustering sensor data)~\cite{van2021event,jablonski2019multi,ruhkamp2021cep}.

\begin{figure}[t]
    \centering
    \includegraphics[width=\textwidth]{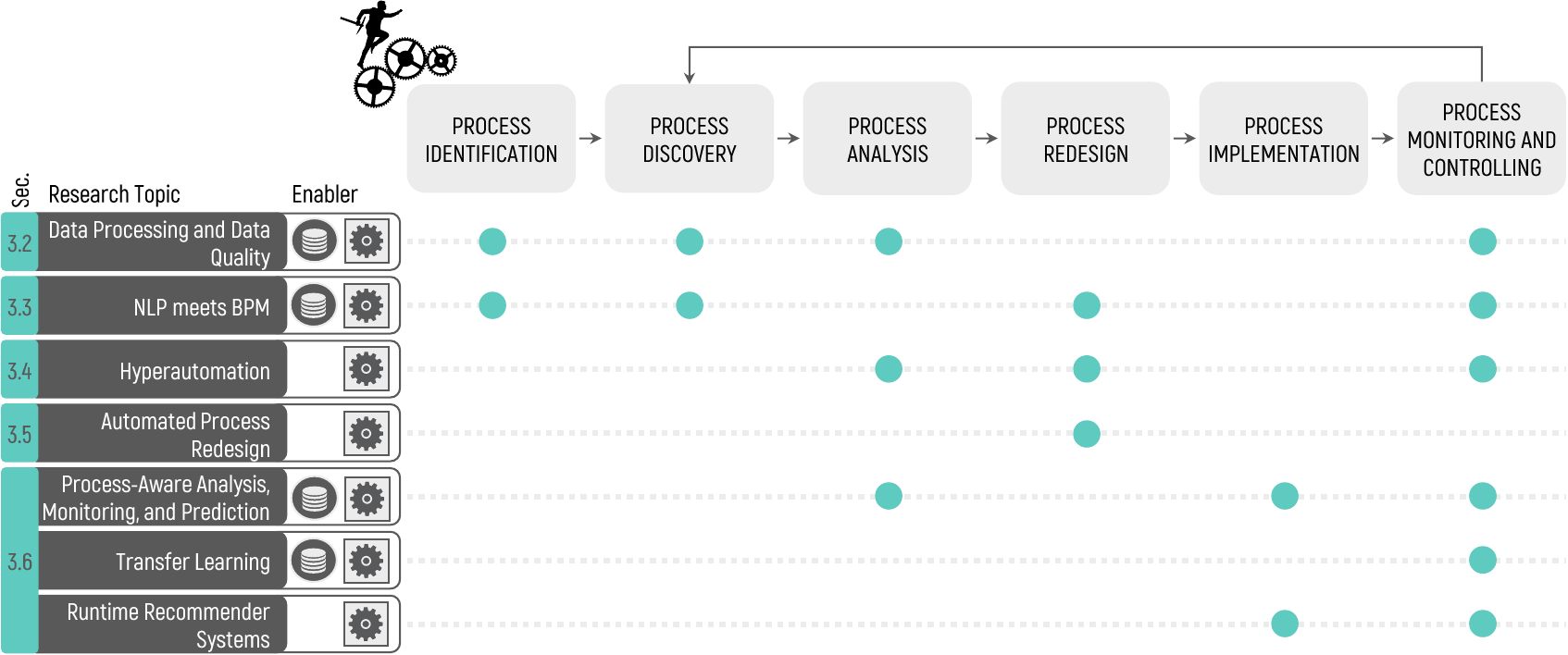}
    \caption{Assignment of research topics (rows) to lifecycle phases (columns); \emph{database} symbol = data as enabler, \emph{gear} symbol = techniques as enabler; Sec. = section discussing the topic as part of a broader research field (see Fig.~\ref{fig:framework-extended} in Appendix~\ref{sec:research_community} for a breakdown of the thematic interests of the research groups involved in this paper).}
    \label{fig:framework}
\end{figure}

Further details of novel techniques and data sources are provided in the subsequent sections discussing five trending research fields that have a high potential to take data-driven \gls{bpm} to a new level and cover the entire process lifecycle. While some of the research fields solely address individual phases of the lifecycle, others have a cross-phase nature. The latter points in particular to a trend towards increased connectivity of individual phases and a more holistic perspective. Fig.~\ref{fig:framework} assigns the individual topics to the respective phases and further indicates whether the topic is mainly focusing on novel data or new techniques and methods. This classification shows a particular focus on the process monitoring phase, although it also indicates that all lifecycle phases are the subject of research in data-driven \gls{bpm}. In addition, the accumulation of research groups in the first row emphasizes the importance of data (pre-)processing and quality. This also confirms the introductory assessment of this paper, according to which the paradigm shift in \gls{bpm} is described as \emph{data-driven}. It can also be deduced that the development of new data sources in current research activities always requires new types of technology (see \emph{Enabler} column).

The subsequent sections are structured according to the process lifecycle phases. Each section first starts with a brief introduction to the research field and motivates its (practical) relevance. Based on the state of research discussed subsequently, the section focuses on an outlook on how the research field will develop in the future.

\subsection{Research Field: Data Processing and Data Quality}\label{sec:data-processing}

For process mining, process-relevant data traces are transformed into standardized event logs, facilitated by predefined \gls{etl} processes. Subsequently, these event logs are leveraged for process mining analyses. 

So far, data for event logs has been obtained primarily from \gls{pais} containing data in a readily applicable structured format and similar granularity, e.g., data available in tables of a relational database of an \gls{erp} system. However, these only represent a small portion of the generated data, potentially leaving a large amount of process information unexploited for process mining analyses~\cite{DBLP:journals/corr/abs-2401-13677}. Many key process activities often remain undocumented in existing systems, limiting their potential for process mining applications. This deficiency stems from the fact that critical process activities often occur outside of information systems, leaving their data traces dispersed across peripheral systems. Given that approximately 80\% of a company's data is generated in unstructured format, a significant portion of process-relevant data goes unused for analysis. This issue is exacerbated by the continuous growth of unstructured data~\cite{balducci2018unstructured}. Examples include data stemming from production systems, manually tracked manufacturing processes, or communication via phone or email, where relevant data may reside in email archives~\cite{di2013mining}, textual documents, sensor logs, images and video recordings~\cite{DBLP:conf/iceis/FichtnerSJ20a}, recovered residual data~\cite{DBLP:conf/ifip8-1/EnglbrechtSP20}, or network traces~\cite{hornsteiner2024,Empl2024}. 

The focus of this research field therefore is on the identification and connection of new data sources for process mining as well as on the processing and correlation of exploited unstructured data and events. Currently, there are several works dealing with different aspects of event correlation~\cite{ferreira2009,koschmider2019}. There are also first approaches that deal with the transformation of data traces available in unstructured formats (e.g., text data~\cite{banziger2018,chambers2020}, time-series~\cite{DBLP:conf/caise/FongerAK23}, sensor data~\cite{hemmer2020,leotta2020}, video data~\cite{knoch2018,KRATSCH2022,DBLP:conf/bpm/LepsienKK23}, or network data~\cite{Coltellese2019,hornsteiner2024,engelberg2021}) into standardized event logs.

However, these works usually focus on a specific use case (e.g., customer relationship management systems~\cite{banziger2018}, job store manufacturing~\cite{knoch2019}, recruitment systems~\cite{engelberg2021}) and a specific type of (un)structured data. Furthermore, there are hardly any approaches that enable the combination of different data sources (both structured data sources among themselves and in combination with unstructured data sources). A generalizable, modular framework that allows for the (semi-)automated use of both structured and unstructured data types from different heterogeneous data streams and systems under consideration of data quality and event correlation does not exist at the moment. Hence, there is a need for a more generally applicable, holistic solution that combines multiple (un)structured data sources and enables various application scenarios.

\begin{figure}[bth]
    \centering
    \includegraphics[width=\textwidth]{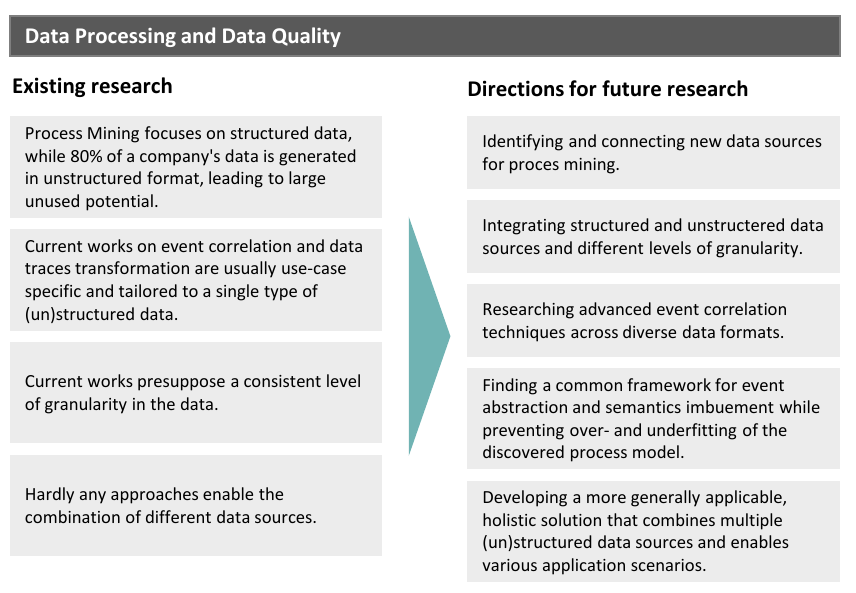}
    \caption{Data Processing and Quality - now and in the future}
    \label{fig:processing}
\end{figure}

Therefore, events and activities from different systems must be correlated to effectively carry out process mining across system boundaries. Typically, data traces from sensor streams or peripheral systems lack structure and may even be entirely unstructured, making it challenging to correlate them with data stemming from other systems. The majority of existing process mining methods presuppose that event data are recorded at a consistent level of granularity. However, data traces originating from different heterogeneous information systems, whether structured or unstructured, often go unaggregated due to a lack of event correlation~\cite{beverungen2021seven}. These systems may track process activities differently, impacting the discovered process model from the logs they generate. This can result in incomplete process evaluations and the omission of actionable data, especially in advanced analytics like predictive process monitoring, which heavily rely on high-quality and comprehensive databases~\cite{beverungen2021seven}.

Two essential steps are required to ensure appropriate data quality and to make raw logged, unstructured data from various systems and data streams suitable for process mining: \emph{(i) pre-processing} and \emph{(ii) event-activity abstraction}. Pre-processing of data stemming from various heterogeneous data streams, sensors, systems, and network traffic is a labor-intensive and error-prone task, encompassing data integration, enhancement, transformation, reduction, discretization, and cleaning. Pre-processing is crucial to transforming raw data into a usable format, removing noise and outliers (e.g., erroneous or missing values), and extracting representative data~\cite{DBLP:conf/bpm/KoschmiderKKZ21}. A comprehensive survey of data pre-processing can be found in~\cite{marin2021event}.

Subsequently, pre-processed data must be imbued with semantics through \emph{aggregation techniques}, elevating the data to a \emph{higher level of abstraction}. Various techniques exist to extract high-level events from this pre-processed data. However, the challenge of abstraction lies in adding semantics to input data that may not be fully understandable. Overly fine-grained abstraction can lead to overfitting in the discovered process model, while overly coarse-grained abstraction results in underfitting. For structured data, a comprehensive survey on event abstraction was conducted in~\cite{van2021event}. A common framework for tackling unstructured data, however, is for the most part an open research problem.

The ability to use (un)structured data from a wide variety of heterogeneous data streams and systems allows all process-relevant data to be exploited for further analysis~\cite{schoenig-network-2022}. Significant value is therefore created by the following new possibilities: \emph{(i)} it enables a comprehensive and exhaustive mapping of processes to event logs, which can then be used for further process analyses and evaluations, \emph{(ii)} it enables an increased log coverage leading to a better end-to-end understanding of processes and related issues based on newly acquired information, \emph{(iii)} it supports the development of a holistic concept for the identification, use, and evaluation of process-relevant data in a process mining context and \emph{(iv)} it enables a joint use of structured and unstructured data for process mining use cases.

In conclusion, bridging the gap between peripheral process activities and structured or unstructured data sources is crucial in the context of process mining. Proper data pre-processing and meaningful event-activity abstraction are essential steps to make raw data suitable for analysis and to ensure that valuable process information is not lost in the complexity of disparate data sources and information systems.

\subsection{Research Field: Process Discovery} 

While process mining so far utilizes structured information from process records (i.e., event logs), an intersection of \gls{nlp} and \gls{bpm} enables utilization of informally represented information sources such as process manuals, interview transcripts, standard operating procedures (SOPs) or process-oriented quality management documentation (e.g., ISO 9001). The genesis of this topic primarily lies within the field of \gls{nlp}. It is closely related to the subfields of information extraction, dedicated to the automated extraction of structured information from unstructured texts in general. However, the ability to extract process-relevant information from texts is essential in several phases of the process lifecycle. Thus, using \gls{nlp} text classification techniques, the (partial) automation of process identification and the classification of processes as either support or core processes holds significant potential for substantially enhancing the efficiency of \gls{bpm} initiatives. 

Furthermore, studies reveal that creating a formal business process model manually can consume a significant portion of the overall time allocated to a \gls{bpm} project, with estimates indicating up to 60\% of the total project time~\cite{Friedrich2011}. \gls{nlp} information extraction techniques have the potential to automate the transformation from informal process descriptions to process models and, thus, to increase the efficiency of both the process discovery and process redesign phases. Thus, while the motivation here is the same as for traditional process discovery on event logs, \gls{nlp} provides access to so far undiscovered and neglected information sources, i.e., process information as natural language texts. This further holds for the execution phase of the business process life cycle, too. This phase typically requires the processing of different types of data, often encompassing unstructured data such as emails, support requests, or notes from process participants. This unstructured textual data can hold valuable information that may affect the proper execution of processes and condition a positive process outcome. To summarize, the application of modern \gls{nlp} techniques is advantageous for at least two reasons: First, processes already documented in text form can be modeled automatically, which enables the application of standard \gls{bpm} techniques (e.g., for execution and monitoring) and, second, data generated in the process in text form can be used automatically (e.g., for the automation of decision logic in process control).

Previous research focused on rule-based approaches for the above-mentioned \gls{nlp} tasks in \gls{bpm} (\cite{Friedrich2011}, etc.). Rule-based approaches have inherent limitations, including low adaptability and high maintenance requirements. However, the evolution towards more powerful \gls{ml}-based approaches is hindered by the lack of \gls{bpm}-specific training datasets. Therefore, new techniques are being developed hand-in-hand with new datasets, which mainly focus on automating the task of creating process models from unstructured information. As of now, no approach has completely resolved this task. However, there are partial solutions available, such as the extraction of information relevant to a business process, including activities, actors, and their relationships~\cite{vanDerAa2019,Ackermann2021,Borges-Ferreira2017,Qian2020,Quishpi2021}. Yet, these approaches often disregard essential parts of the task, such as structuring and pre-processing the input text~\cite{Qian2020}, extracting relationships between different process elements~\cite{Ackermann2021}, or generating the complete business process model~\cite{Borges-Ferreira2017,Neuberger2023}. In addition, a lack of data prevents the development of approaches based on modern deep learning models, which are otherwise generally established for processing natural language information. Currently, the largest dataset available (PET~\cite{Bellan2022}) comprises only 47 process descriptions. Hence, a vast number of approaches resort to rule-based methods, as the available data is insufficient for the successful application of \gls{dl} models. Consequently, our own research focuses on techniques for improving the data basis first (e.g., using data augmentation, oversampling, and supporting the manual annotation task for creating datasets with \gls{dl}). At the same time, we are researching techniques that contribute to the holistic extraction of process models from natural language texts (e.g.,~\cite{Ackermann2023}). In order to cover both scenarios with a small amount of available data and those with an extensive database, our research includes two basic principles: i) the development of \gls{ml} pipelines~\cite{Neuberger2023} and ii) the application and fine-tuning of \gls{dl} and/or (large) language models~\cite{Ackermann2021}. 

\begin{figure}[bth]
    \centering
    \includegraphics[width=\textwidth]{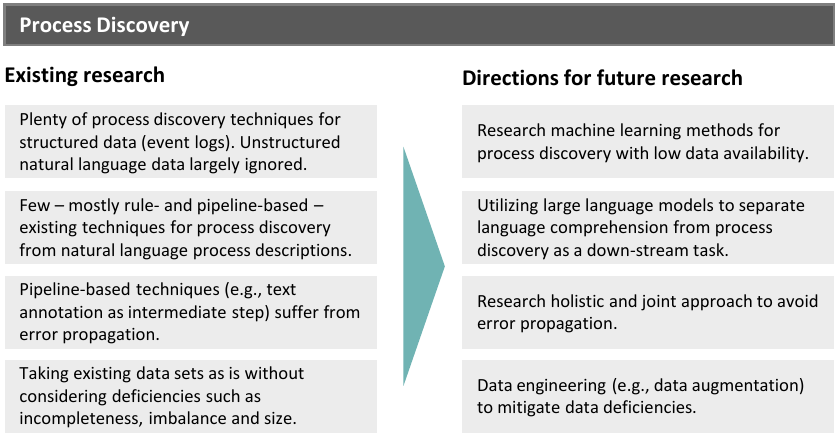}
    \caption{Process discovery - now and in the future}
    \label{fig:process_discovery-overview}
\end{figure}

In all the above-mentioned use cases, the rapidly advancing development of pre-trained (large) language models (e.g., BERT, ChatGPT) offers significant potential~\cite{devlin2018bert}. These models provide foundational capabilities for comprehending natural language, making them valuable building bricks that can be fine-tuned to address tasks effectively and holistically. Thus, future work in this research field needs to investigate a paradigm shift from training isolated pipeline components providing partial solutions of this task (e.g., extracting activity labels or matching multiple mentions of the same process entity) to tuning queries to \gls{llm}s able to solve multiple of those tasks without being explicitly trained to do so. However, in various application domains, it is still beneficial to fine-tune \gls{llm}s to specific tasks rather than solely relying on their pre-trained knowledge. Thus, research in this area will also need to investigate fine-tuning for \gls{llm}s for extracting process information from natural language text. Additionally, the generative capabilities of \gls{llm}s could also be used in a preparatory step to improve the data basis itself (e.g. for data augmentation). The paradigm shift of using \gls{llm}s raises novel challenges both for the extraction of process information from texts and for the use to improve the data basis. These include the partitioning of inputs based on input text length restrictions, the handling of output length restrictions, the formatting of input and output messages, and the need for specialized input formulations (\emph{prompt engineering}). However, the vision is already very tangible: In the near future, process analysts will conduct process interviews with a smartphone on the table and a process model will be created fully automatically in the background based on this interview.

\subsection{Research Field: Hyperautomation}
\begin{figure}[bth]
    \centering
    \includegraphics[width=\textwidth]{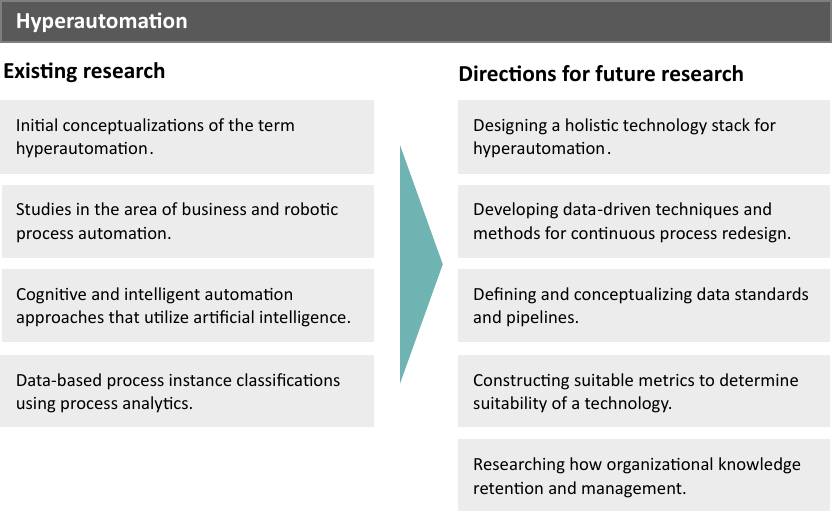}
    \caption{Hyperautomation - now and in the future}
    \label{fig:hyperautomation-overview}
\end{figure}
The global competition, inflation of resource prices, and shortages in skilled personnel force companies to become more efficient in their operations.
Business process automation represents a potential solution to such challenges as it can, for instance, increase human resource utilization and decrease cost and processing times~\cite{denagama2020empirically}. 
Therefore, companies automate their processes to gain competitive advantages in their markets~\cite{aguirre2017automation,denagama2020empirically}.
However, the continuous advancement of technology results in shorter intervals for evaluating whether tasks can be automated or rely on human execution~\cite{aalst2018robotic}.
Thus, companies are in a constant search for opportunities in the area of automation to stay competitive. This constant and overarching drive towards automation results in the phenomenon of hyperautomation~\cite{xiaohui2022hyper}. Hyperautomation combines \gls{bpm}, \gls{ai}, \gls{rpa} and other technologies to maximize automation potential~\cite{haleem_hyperautomation_2021,jimenez-ramirez2021journeyhyper}. The focus of hyperautomation moves beyond automating tasks towards automating entire business processes~\cite{lasso2020hyperautomation}. Applying hyperautomation to \gls{bpm} helps organizations to achieve a holistic process landscape. 
By harnessing synergies between different technologies, hyperautomation enables data-driven identification of automation potential through analysis.
Although hyperautomation has an impact on the entire \gls{bpm} lifecycle, it primarily impacts how companies analyze and redesign their business processes. 
Additionally, companies require means to monitor and control their business processes regarding their automation to leverage continuous improvement capabilities. 


%
%
%


Beyond task and process automation, hyperautomation strives for continuous automation efforts~\cite{jimenez-ramirez2021journeyhyper}. 
Regarding complexity, it thereby ranges from basic to cognitive automation. Moreover, hyperautomation initiatives orchestrate various automation approaches on task, process, operational, and organizational levels.
However, selecting the right automation technology poses great challenges to organizations, and making incorrect choices can have detrimental effects~\cite{engel_cognitive_2022,lacity_becoming_2021}.

Understanding \emph{what} to automate and \emph{how} to monitor is a vital aspect of hyperautomation and connects it to business processes and their management~\cite{aalst2018robotic}. 
\gls{bpm} and its data-driven subdomains---e.g., process mining---leverage identifying automation potentials and monitoring automation degrees~\cite{jimenez-ramirez2019earlyRPA,wellmann2020framework}. 
In turn, hyperautomation facilitates structuring the roadmap towards cognitive automation from the business and \gls{it} levels. 
Due to its multidisciplinary nature, the technical hyperautomation stack needs to be managed at the organizational level to succeed on the adoption journey and mitigate failure risks. Hence, one potential solution to the automation challenges posed by organizations is the implementation of hyperautomation. 
%

Due to the infancy of the field, there are knowledge gaps that must be addressed to initiate the research on hyperautomation~\cite{herm2021symbolic,jimenez-ramirez2021journeyhyper}.
A multilateral approach is required to derive characteristics in a data-driven manner and create a set of rules that consider non-technical requirements~\cite{axmann2021framework,herm2021symbolic,mendling_how_2018}. 
%
Because hyperautomation combines different more mature technologies, it can draw from existing knowledge in technology areas such as \gls{rpa}~\cite{axmann2021framework,herm2023framework}, intelligent automation~\cite{herm2021symbolic}, desktop activity mining~\cite{wanner_process_2019}, process mining~\cite{fischer_composition_2021,wellmann2020framework}, and process prediction~\cite{heinrich2021process}. \gls{bpm}~\cite{fischer_strategy_2020} provides a sound starting point for this area of research. 
%
%
%
%
%
%
%
%
%
%
Building on the technology stack, future research could investigate the following challenges from a technical and sociotechnical perspective.

First, adaptable technology stacks for hyperautomation projects in different organizations are to be derived. 
Second, to succeed in the long term, organizations must rely on data-driven techniques and methods to continuously rethink business processes to adjust to technological advances. Future research could guide the introduction and adoption of continuous rethinking by providing the required tools and approaches.
Third, when organizations want to automate a large share of their processes, they must proceed in their business's digital transformation. Meanwhile, to enable data-driven monitoring of automation and human work, process data is recorded. More and more potentially sensitive data is stored in computers, clouds, and databases. Therefore, developing data standards and processing pipelines that comply with the European \gls{gdpr} is a critical field to drive hyperautomation onwards.
Fourth, appropriate \glspl{kpi} must be identified which enable organizations to assess their automation levels or to determine the suitability of a technology for a certain task.
Fifth, hyperautomating an organization's operations comes at the risk of losing knowledge and workers in the core area of business. Hence, future research could investigate how organizations can manage knowledge retention and human-resource reallocation to business areas where human workers thrive. 
Lastly, guidelines and maturity models for technology risk management, automation liabilities, and strategic agility within organizations are subjects of future research.

\subsection{Research Field: Automated Process Redesign} %

In the light of multiple simultaneous crises and rising expectations from both customers and employees, organizations need to continuously adapt their processes through process redesign, i.e., \gls{pii} ~\cite{vanDun.2023,Kerpedzhiev.2021,KreuzerDSS,Malinova.2022,ParkVanderAalst.2022}. Hence, process redesign as a key value-adding activity within the \gls{bpm} lifecycle needs to accelerate so that organizations can sustain or create competitive advantage ~\cite{vanDun.2023,Zellner.2011}. Thereby, process redesign includes both process improvement and process innovation, combining an exploitative and an exploratory approach. However, this requires a high number of resources, both financially and in terms of time, making it expensive and tedious ~\cite{dumas2018fundamentals,Gross.2019,Malinova.2022,vanDun.2023}. Computational support and automation are therefore crucial for organizations’ \gls{pii} activities ~\cite{Beerepoot.2023}. Thus, \gls{piis} which support the generation, evaluation, and selection of improved process designs emerge as an upcoming research area within \gls{bpm}.

This emerging class of systems faces many challenges in this regard ~\cite{Röglinger.2021}: \gls{pii} is knowledge-intensive and highly context-sensitive – domain expertise must be taken into account but is often intangible. In addition, there are numerous restrictions in terms of feasibility and economic viability. For example, dependencies within and between several processes must be considered while certain goals and performance criteria must be met. Furthermore, \gls{pii} requires creativity to come up with innovative solutions ~\cite{Figl.2016}. 
Research can exploit the potential of computational support if it achieves to adequately address these challenges, bridging the gap from insights towards action and taking process analytics to the next level towards value generation in prescriptive \gls{bpm} through automating \gls{pii}. This exhibits the potential to make process design and redesign more effective through utilizing knowledge inherent in the data (e.g., regarding process deviance ~\cite{vanDun.2023}). \gls{piis} can also help to make \gls{pii} activities faster, more efficient, and less expensive as well as less dependent on skilled human process designers' capacity and creativity, thereby overcoming potential resource bottlenecks. Hence, research on the design, implementation, and use of \gls{piis} in practice is relevant for both academia and industry, as it allows for more effective and more efficient process design and evaluation. 

Building on the potential presented by the advance in \gls{genai} and the increasing availability of process execution data, there is already an increasing volume of research on \gls{piis}. In most cases, research has followed a design science or software engineering approach. So far, research has focused on investigating the potential of certain technological approaches such as rule- and heuristic-based approaches (e.g., our research on assisted business process redesign ~\cite{Fehrer.2022}) or evolutionary algorithms (e.g., ~\cite{Afflerbach.2017}), but also approaches using \gls{nlp} (e.g., ~\cite{Mustansir.2022}), and data mining (e.g., ~\cite{Truong.2016}). There are also a few initial approaches leveraging the potential of \gls{genai} for computational creativity (e.g., our own research supporting human process designers via generative adversarial networks ~\cite{vanDun.2023}) or \glspl{llm} (e.g., ~\cite{Beheshti.2023}). 

So far, the few computational approaches to \gls{pii} have predominantly explored technical feasibility in a rather isolated and incremental manner, mostly neglecting the potential inherent in process execution data. Thus, further research on this emerging class of systems fosters data-driven process design generation. The breakthrough lies in the utilization of additional data (both internal and external) as well as in applying state-of-the-art technology for ideation, evaluation, and selection. Especially in the context of \gls{genai}, there is high, yet unexplored potential. For instance, there is hardly any holistic approach combining several activities (i.e., from idea generation to selection) or several technological approaches (e.g., combining \gls{genai}, planning algorithms, and pattern-based approaches). Moreover, there are more data sources to be explored, for example, outside of organizations’ own process data. Moreover, moving from a single-artifact view towards a broader perspective, there still is a need for descriptive and prescriptive design knowledge regarding \gls{piis}, e.g., through examining users’ preferences. Consequently, in our research, we currently focus on both technical design knowledge on a single-artifact level, accompanied by situative implementations (e.g., utilizing LLMs for \gls{pii}), and a more general perspective on the whole class of systems, e.g., investigating current and prospective functionalities, system architectures, and general guidelines, toward a nascent design theory for \gls{piis}. To ensure practical relevance in our research, we closely collaborate with industry partners, for instance, in our publicly funded research project \emph{Next Best Process}.

\begin{figure}[!htb]
    \centering
    \includegraphics[width=\textwidth]{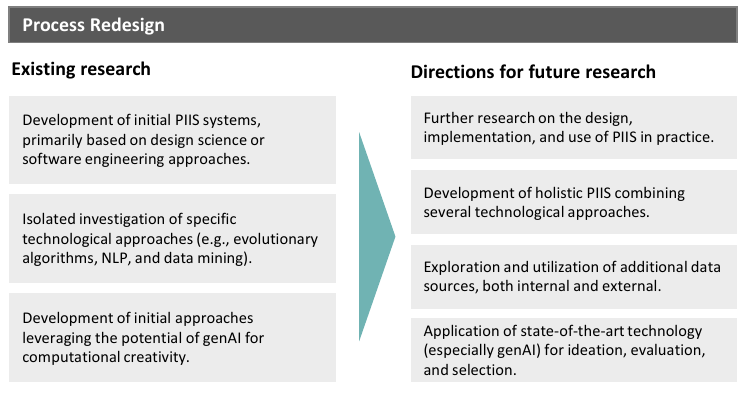}
    \caption{Process redesign - now and in the future}
    \label{fig:process-redesign-overview}
\end{figure}

\subsection{Research Field: Predictive and Prescriptive Business Process Monitoring}\label{sec:pbpm}
The process monitoring phase of the process lifecycle proactively supports process users or other related stakeholders in their decision-making ~\cite{Marquez2017}. \emph{Predictive business process monitoring} is an area of research in this phase that has become increasingly important in recent years~\cite{DiFrancescomarino2022}. In contrast to descriptive and diagnostic analysis of event data that focus on the past, predictive business process monitoring aims to predict how a running process instance will unfold up to its completion including predicting behavior-related (e.g, next steps), process-outcome-related (e.g, expected result and performance), or time-related properties (e.g., remaining time) ~\cite{Verenich2019} and the monitoring of (complex) compliance requirements (e.g., regulatory documents in the financial domain). Process prediction models are built on records of already completed process executions (of the same process)~\cite{DiFrancescomarino2022}. Knowing the further course of the process in advance enables process participants involved in its execution to identify potential problems at an early stage so that preventive measures can be taken in time~\cite{DiFrancescomarino2022,Teinemaa2018}. Also, in case of compliance violations, actionable mitigation measures can be enforced automatically.  \emph{Prescriptive business process monitoring} therefore builds upon predictive process monitoring to realize the value of predictions through actions, mostly through recommending interventions during runtime~\cite{Kubrak2022}. Both predictive and prescriptive business process monitoring exhibit potential for research.

The first predictive approaches in \gls{bpm} were developed in the middle of the last decade and relied on an explicit representation of the underlying process model (e.g., probabilistic finite automata~\cite{Breuker2016} or hidden Markov models ~\cite{Lakshmanan2015,Unuvar2016}). With the increasing popularity of \gls{dl}, various \gls{dl} architectures have been used for various tasks in process monitoring~\cite{Rama-Maneiro2023}. Unlike the pioneer approaches, they have the advantage that they no longer require a process model of the process and, hence, are completely data-driven. Several studies show the superiority of \gls{dl} approaches~\cite{Kratsch2021,Verenich2019}, which is why they have become state-of-the-art in predictive business process monitoring~\cite{Rama-Maneiro2023}. However, they have several shortcomings:
\begin{itemize}
    \item \emph{Risk of replicating the past}: Limited solution space based on historical data can result in undesirable recommendations based on past actions~\cite{Weinzierl2020}.
    \item \emph{No inclusion of domain knowledge}: Business processes are not executed in isolation. They are always bounded by their surroundings, that is, the industry the organization operates in, the required output (product or service), the resources necessary, regulations to be adhered to, as well as potential risks associated~\cite{Weinzierl2020}.
    \item \emph{Limited adaptability}: 
    Business processes are inherently unique and tailored to a company or organization and evolve due to factors like innovation, experience, regulations, crises, and rising expectations. This dynamic nature demands flexible adaptability of prediction models. Currently, models must be trained anew for each process, requiring retraining from scratch with any process changes.
    \item \emph{Data hungriness}: 
    Training predictive models requires substantial hardware resources and time, with the quality and performance heavily reliant on having sufficient quantitative and qualitative amounts of diverse data~\cite{Kaeppel2021a,Kaeppel2023}. However, this requisite is often not met due to long process runtimes, infrequent executions, or seldom-executed processes~\cite{Kaeppel2021b}. As a result,  prediction models are either not trained well enough to make profound predictions (especially in special or failure cases) or only cover cases explicitly present in the training data~\cite{Kaeppel2021c}.
    \item \emph{Inefficient training}: 
    Training inefficiency arises from low variance in event logs, resulting in redundant learning of similar process instances and behaviors~\cite{Kaeppel2021a}. Since for each business process, a separate prediction model is trained, process behavior that is shared by multiple processes must also be learned several times. 
    \item  \emph{Lack of transparency}: \gls{dl} models largely represent a black box, i.e., their decision-making process cannot be understood in detail caused by the large number of parameters (i.e., trainable weights)~\cite{Nauta2023}.
    \item \emph{Limited automation capabilities:} 
    Currently, process mining is used to uncover process models, identify tasks for automation, and enable ongoing monitoring post-automation (so-called \emph{mine and automate} pipeline). However, it has several drawbacks~\cite{RinderleMa2023a}:
    \begin{itemize}
        \item \gls{rpa} automates only single, simple, and repetitive interactions of humans with software. However, that is insufficient, as processes are task-overarching and orchestrating concepts.
        \item It is restricted to available data sources (often collected for other purposes) rather than (pro-)active collection of contextualized data.
        \item 
        It analyzes processes retrospectively instead of during runtime, limiting the handling of uncertainty, drifts, and exceptions.
    \end{itemize}
    \item \emph{Including multiple compliance requirements:} 
    Multiple compliance requirements are currently incorporated as prediction goals into a \emph{comply and predict pipeline}, where each requirement requires its own prediction~\cite{RinderleMa2023a}.
\end{itemize}

To overcome these challenges, runtime recommender systems, transfer learning, and process-aware automation offer considerable potential. 

\paragraph{Runtime Recommender Systems}
To overcome the risk of replicating the past and to include domain knowledge, approaches from the field of neuro-symbolic \gls{ai} can be applied to business processes~\cite{Weinzierl2020}. Neuro-symbolic \gls{ai} aims at combining the strength of neural networks and symbolic \gls{ai} that rely on logic, rules, and symbolic representations to create effective and \gls{xai}-based systems~\cite{Sarker2021}. This combination enables prediction models to provide recommendations instead of predictions. These recommendations (e.g., the next best action) of so-called \emph{runtime recommender systems} can include explanations and can be tailored to process users or other related stakeholders~\cite{branchi2022learning}. 

Utilizing neuro-symbolic \gls{ai} is essential for combining knowledge and cognition~\cite{sheth2023neurosymbolic}. This is required for creating meaningful runtime recommendations and to provide explanations on why a specific recommendation is given~\cite{townsend2019extracting}. This integration offers several benefits: First, recommendations based on domain knowledge lead to more satisfactory prescription results, enabling a process user or other related stakeholder to enhance business process performance. Second, it makes implicit knowledge explicit, allowing process users or other stakeholders to comprehend how the recommender system creates its recommendations and, therefore, give them greater consideration to recommendations from the system. Third, it reduces the required amount of event data to train models for predictive and prescriptive business process monitoring, making it more accessible to organizations that do not have enough data (cf. data hungriness).

Relevant domain knowledge about business processes exists in various forms (e.g., discovered process models, traditionally modeled process models, or gained insights from conformance checking). To empower \gls{dl}-based approaches with such knowledge, \gls{bpm} and process mining research must investigate how and to which extent such domain knowledge can be embedded into ontologies, knowledge graphs, or simply encoded into process data.

The primary objective of such novel runtime recommender systems is to prevent the replication of common and undesirable behavior within a business process~\cite{weinzierl2020detecting}. To achieve this, the integration of data that allows \gls{ai}-based, specifically \gls{dl}-based approaches, to learn from further data captured in the entire body of knowledge in the application or \gls{bpm} domain is essential. Reproducing past behavior also inhibits the establishment of improvement targets, for instance, regarding \glspl{kpi} or compliance measures. 

Process-aware execution systems usually grant a certain degree of freedom to process users and other related stakeholders so that they can seamlessly perform their tasks~\cite{dunzer2023design,jans2013case}. This freedom during execution can be utilized to find more efficient and effective execution variants~\cite{weinzierl2020detecting}. However, \gls{dl}-based techniques used in recommender systems must be able to distinguish allowed and prohibited process behavior. For instance, physical laws may prevent applying a coating to a material earlier, general logic prevents assembling a final product before its parts are produced, or compliance rules forbid the release of a payment before it is approved~\cite{kluza2017method}. This distinction between allowed and prohibited process behavior can be significantly supported by the above-mentioned integration of domain knowledge~\cite{weinzierl_deviation2021}.

In summary, runtime recommender systems offer practical benefits by providing suitable recommendations to users, aiding in compliance, enhancing process performance, and optimizing resource efficiency~\cite{park_performance_2020}. Besides that, they might have the potential to lead to process redesign and improvement. While application scenarios for predictive and prescriptive process monitoring generally focus on the instance level of a business process, it is worth noting that process-level predictions or prescriptions can also hold significant importance for improving the performance of business processes. In this sense, self-adaptive business process execution engines can be built that automatically respond and adapt to changes in the surrounding environment of a business process. Implementing such systems requires attention to technical aspects like integration into execution systems and access to running process data and domain knowledge.

\begin{figure}[bth]
    \centering
    \includegraphics[width=\textwidth]{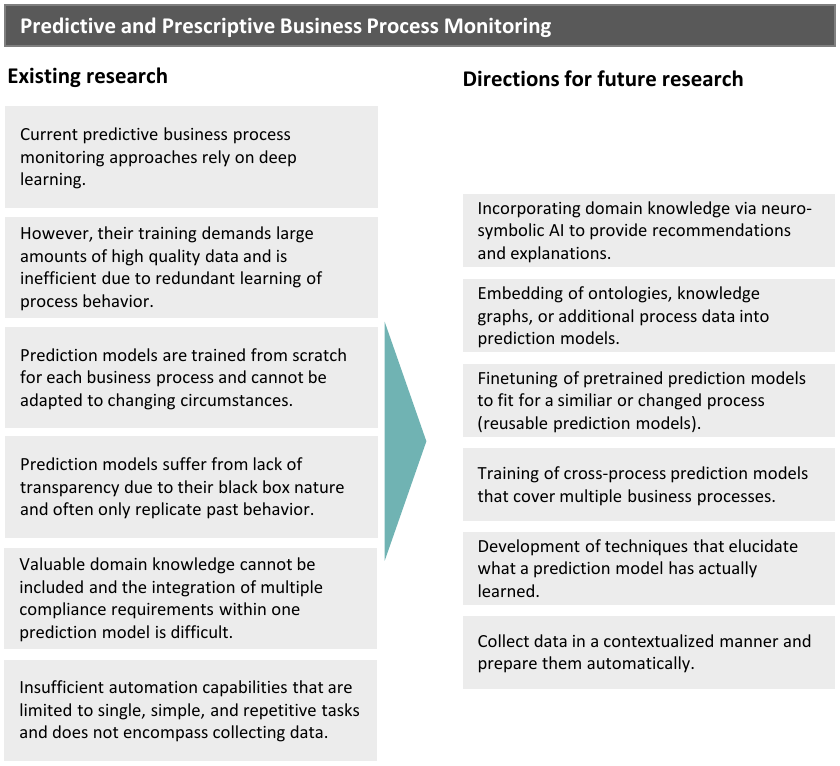}
    \caption{Predictive and prescriptive business process monitoring - now and in the future}
    \label{fig:pbpm-overview}
\end{figure}

\paragraph{Transfer Learning}
In other research fields, attempts are made to solve limited adaptability, data hungriness, and inefficient training of \gls{ml} models by so-called \emph{transfer learning}. In transfer learning, an already trained \gls{ml} model is adapted to a new task or applied to a novel but sufficiently similar task. Empirical research has demonstrated that in transfer learning, the initial quality of the model at the beginning of the training is already higher than in a conventional training process~\cite{Zhao2023}. As a result, the model makes faster learning progress and achieves, in the end, significantly better performance~\cite{Zhao2023}. This is made possible by starting with a model that has already been sufficiently well trained for a similar task or comes with a basic understanding of the task to be learned~\cite{Zhao2023}. 

One consequence of transfer learning is that significantly less training data is required. 
For example, consider two companies, A and B, with an Order-to-Cash process. Both processes exhibit significant similarities and share the same objectives. However, there are also notable differences in certain parts of the process. While Company A executes the process frequently, Company B's execution frequency is notably lower. Consequently, the available data in Company B is insufficient to train a robust and reliable prediction model. Here, transfer learning could be employed, using the model trained for Company A as a starting point and fine-tuning it with Company B's available data to capture its specific nuances. Since fine-tuning generally requires less data, Company B's available data volume is adequate.

Particularly within the domains of \gls{nlp} and computer vision, transfer learning has already found successful applications. For instance, in \gls{nlp}, powerful language models are trained, possessing a general understanding of language, and only need to be adapted to a specific task to be solved~\cite{Griesshaber2020}. Overall, transfer learning is considered a promising research field in \gls{ai}, characterized by substantial innovation potential. Hence, there is a need to advance transfer learning in \gls{bpm} and develop BPM-specific transfer learning approaches.

Overall, transfer learning has the potential to enhance predictive business process monitoring by boosting model performance, enabling cross-process prediction models, speeding up training significantly, and offering customizable and reusable prediction models. The latter points also promote sustainability by saving resources in the training phase. However, these benefits are accompanied by several challenges: First, there is to investigate whether there exists a general understanding of processes that is shared by all business processes (or at least by specific types). Besides such a general "process thinking" the transfer of prior knowledge or contextual knowledge must be enabled. Second, approaches must be developed to check whether processes selected for transfer are sufficiently similar to transfer knowledge from one process to the other in a senseful manner. Additionally, by applying transfer learning between different processes, a legally secure transfer of knowledge between processes must be ensured. This is especially evident when prediction models are transferred across company boundaries, ensuring the preservation of company secrets. A closely related issue, akin to the lack of transparency, is the necessity to develop techniques that elucidate what a model has actually learned. This is essential for assessing the success and scope of knowledge transfer. To date, a hypothesis is that predictive business process monitoring approaches for predicting the next activity implicitly learn a process model based on the event data \cite{Evermann2017}. However, more detailed investigations are still largely in their infancy \cite{Peeperkorn2023}. A visualization of the process knowledge learned by a \gls{dl} prediction model, e.g., in the form of a process model or a similar well-understood representation, would allow process participants and domain experts to easily validate the model and evaluate its potential.

So far, previous research has predominantly focused on training prediction models for each process from the ground up. With the successful application of transfer learning an entirely new learning paradigm would be introduced and would represent a significant step towards equipping \gls{ai} with a universal understanding of processes, as required for a multitude of \gls{bpm} tasks, including process redesign, process optimization, and process automation.

\paragraph{Process-Aware Automation, Mining, and Prediction}
The automation of process orchestrations and choreographies makes an \gls{etl} step in the process mining pipeline obsolete, since data and event streams can be collected in a contextualized manner and are prepared automatically.

The contextualized collection of event and context data requires the development and application of new analysis and prediction techniques~\cite{Ehrendorfer2021,Stertz2020}, that enable the derivation of novel insights to create business value. Techniques for analyzing contextualized data partly rely on conventional \gls{ml} techniques but are configured and applied in \gls{bpm} specific context. Hence, process automation goes far beyond \gls{rpa} and task automation as it supports entire process orchestrations. Another major innovation is the use of automation to collect data \footnote{An example use case for process-aware automation from the manufacturing domain can be found at \url{https://lehre.bpm.in.tum.de/~emangler/.Slides/media/media1.mp4}}. In order to fully benefit from contextualized data collection, additionally, sensor streams can be connected to the process models~\cite{Mangler2023}. 

Another task of process analysis and management is compliance management, including the extraction, preparation, and verification of compliance requirements over process models and instances. The compliance requirements typically stem from textual sources such as regulatory requirements, paving the way to the application of \gls{nlp} and \gls{llm} techniques, see, e.g.,~\cite{Barrientos2023}. Their verification is conducted based on the process models applying model-checking techniques, over process event logs (ex post), and process event streams (online). A new avenue is posed by predictive compliance monitoring where the compliance of processes is predicted at runtime~\cite{RinderleMa2023b}.  

To include multiple compliance requirements, we advocate to reverse the \emph{comply and predict} pipeline into a \emph{predict and comply} pipeline where predictive business process monitoring is in place - and can also be used for predicting other \glspl{kpi} — and the prediction results are used to monitor the compliance states for the requirements. This results in several advantages such as maintainability, flexibility, performance, and transparency. The suggested \emph{predict and comply} pipeline is — from a system perspective — covered by predictive compliance monitoring~\cite{RinderleMa2023b}. Note that in a predictive compliance monitoring system, both pipelines might be followed if, for example, some compliance requirements are of high importance and need to be monitored at any given time. 

The output eventually comprises a collection of contextualized data sets, a set of novel process analysis and prediction techniques, and continuous reports on (predicted) compliance states and their visualizations. Process-aware automation offers many advantages but requires process-oriented thinking and automation (beyond task automation), instead of implementing the \gls{etl} process. Hence, a major objective is to find ways to foster process orientation and automation in order to realize the vision of process-aware automation. Predictive compliance monitoring, though, can be realized as an extension of predictive business process monitoring and, hence, does not necessarily require (full) process automation. 

\section{Outlook and Call to Action}
\label{sec:outlook}

Emerging technologies, as well as new data sources, open up avenues to address innovative use cases and have led to a paradigm shift towards data-driven BPM. This position paper presents five selected research fields spread across the whole process lifecycle, encompassing different research topics we currently face in BPM research. These topics are the primary focus of the authors of the paper at hand and have been discussed and refined in two summits. Without claiming to be exhaustive, these are topics to which the authors attribute a particularly high innovation potential.

Each research topic is motivated by describing why it is worth to be investigated together with outlining its usefulness in the future. We deliberate on challenges to surmount and provide initial thoughts, as well as opening points for addressing them. In summary: We expect to extrapolate new structured data (e.g., network data) and unstructured data (e.g., e-mails, videos, audio) to strengthen the data foundations of business processes and facilitate analytics in entirely new areas of application. On the one hand, leveraging more and various data sources will allow strengthened evidence-based decision-making and efficient process optimization. On the other hand, this also requires suitable technologies to utilize this rich data potential. Due to the great and rapidly accelerating progress in the field of artificial intelligence in very different areas, a large number of such techniques are available. However, the application of these techniques in the \gls{bpm} domain harbors a number of specific challenges. To provide more robust techniques for artificial intelligence in \gls{bpm}, we propose the investigation of different promising machine learning concepts (e.g., transfer learning, federated learning, neuro-symbolic artificial intelligence) that could have a positive impact on data-driven \gls{bpm}. Fostering conceptual and technical support for evidence-based decisions for process improvement and innovation initiatives, automation projects, and reliable data quality is an objective of our joint efforts. Therefore, we expect to elaborate frameworks for reliable data-driven automation in various phases of \gls{bpm}, to develop techniques to measure and improve process data quality, and we propose \gls{ai}-based recommendations for implementing business processes.

We hope that this overview has outlined the manifold opportunities data-driven \gls{bpm} offers and will serve as inspiration for future research for both novice and advanced researchers. At the end of this paper, we would like to explicitly call for collaboration and encourage you to take up the ideas presented in your research to start the next generation of \gls{bpm}. We also encourage you to critically discuss the ideas presented and expand them to include other topics and trends.

\subsection*{Acknowledgements}
This project has received funding from baiosphere - the bavarian ai network. 

%
%
%
\bibliographystyle{abbrv}
\bibliography{references}
\appendix

\section{Research Community}
\label{sec:research_community}
\begin{figure}[ht]
    \centering
    \includegraphics[width=\textwidth]{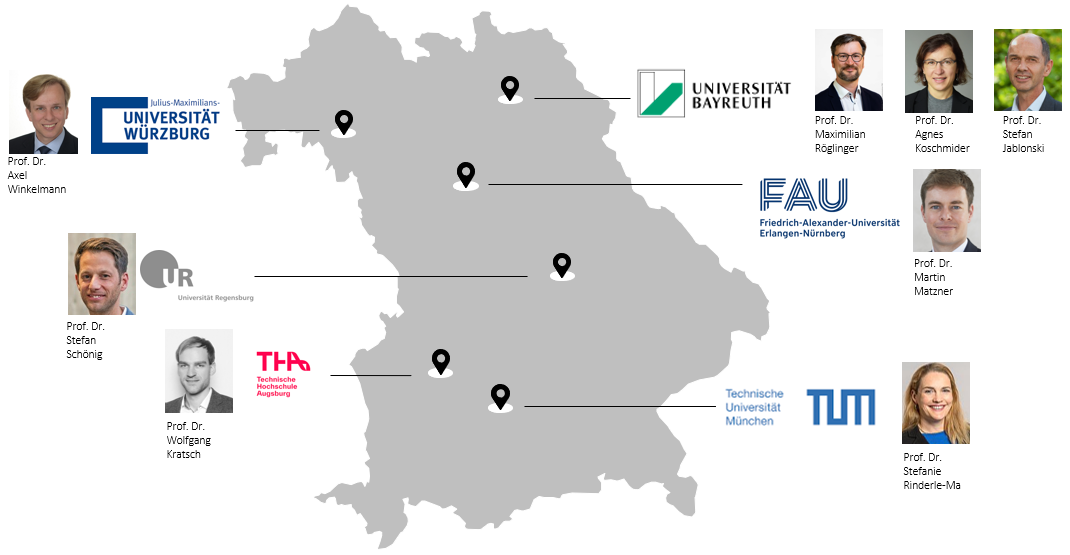}
    \caption{Overview of research groups in the research network}
    \label{fig:chairs}
\end{figure}

\begin{figure}[ht]
    \centering
    \includegraphics[width=\textwidth]{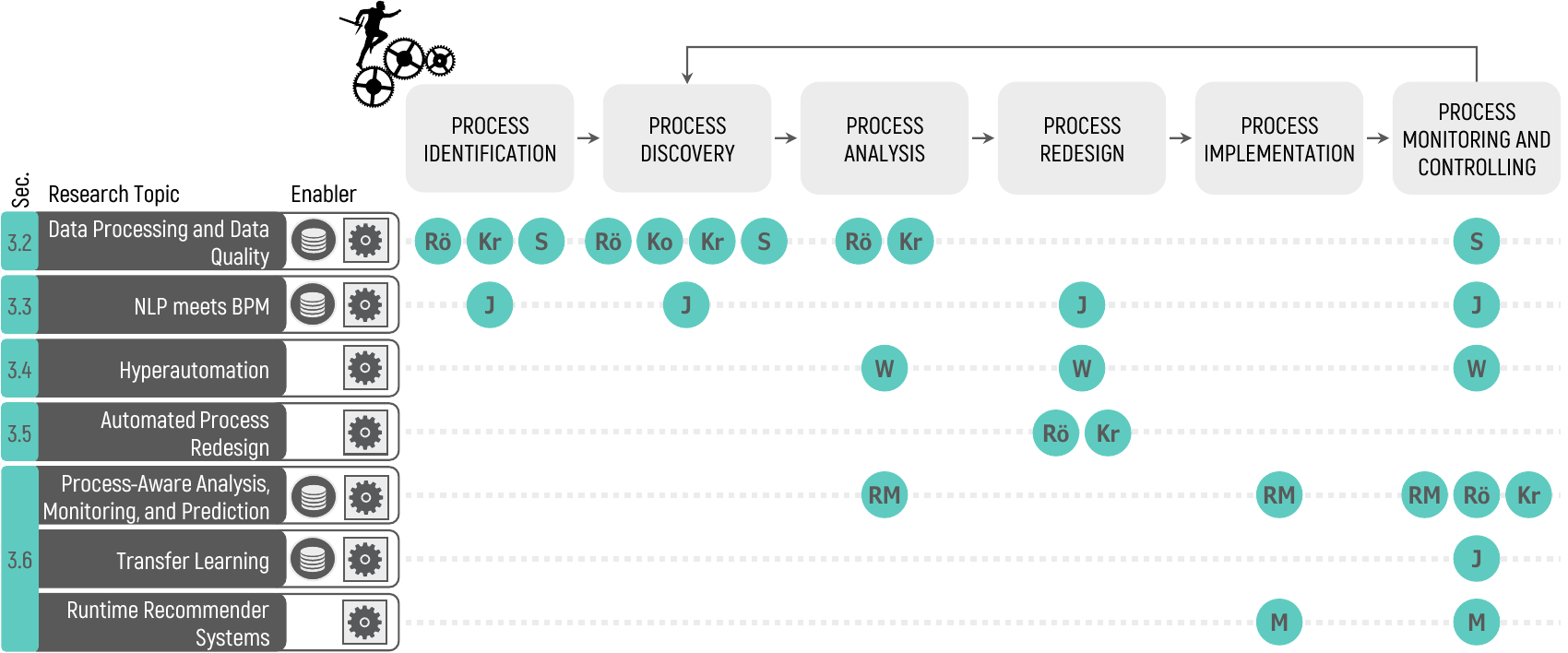}
    \caption{Assignment of research topics (rows) and research groups (cells) to lifecycle phases (columns); \emph{database} symbol = data as enabler, \emph{gear} symbol = techniques as enabler; Sec. = section discussing the topic as part of a broader research field.}
    \label{fig:framework-extended}
\end{figure}

The \gls{dproma} research network brings together experts from six different Bavarian universities. All members of the network contribute to this paper with complementary knowledge in the field of \gls{bpm}, combining computer science, information systems, and management.
Fig.~\ref{fig:chairs} provides an overview of the research groups in the research network and the subsequent list provides a more detailed account of their main research interests. Further, Fig.~\ref{fig:framework-extended} outlines the assignment of research topics and research groups to lifecycle phases. We added a unique \textcolor[HTML]{5fcbc0}{green badge} to the subsequent description of the research groups to link them directly to the topics in the figure.

Prof. Dr. Axel \textbf{Winkelmann}\chairmarker{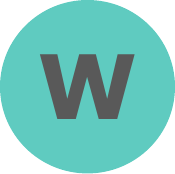} holds the \emph{Chair of Business Ma\-nage\-ment and Business Information Systems} at the Julius-Maximilians-University of W\"urz\-burg focusing on enterprise computing for enterprise systems. His research encompasses areas like \gls{bpm}, hyperautomation, and \gls{ai} integration. Emphasis is on process mining, \gls{rpa}, \gls{ai}-based decision support, especially \gls{xai}, and the role of digital platforms and decentralized ledger technology in value networks, addressing both opportunities and challenges. The team emphasizes socio-technical systems' potential, with projects centering on sustainability, work design, and digitalization. Through hyperautomation research, they enhance automation for small and medium-sized enterprises, boosting resilience and efficiency.

Prof. Dr. Stefan \textbf{Sch\"onig}\chairmarker{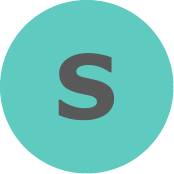} is head of the research group \emph{IoT-based Information Systems} at the University of Regensburg. His teaching, research, and various practical projects address \gls{bpm} and the \gls{iot} in general and particularly focus on the industrial environment, such as an \gls{iot}-based location management tool for the crafts sector in collaboration with practice partners and investigates \gls{bpm} as a driver for cybersecurity in the \gls{iiot}. Prof. Dr. Schönig and his team are focusing on data availability and quality issues for process mining, e.g., processing and using \gls{iiot} network data for process mining, as well as developing process discovery techniques particularly designed for IIoT network data.

Prof. Dr. Wolfgang \textbf{Kratsch}\chairmarker{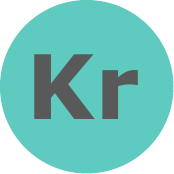} is a professor of \emph{Applied \gls{ai}} at the Technical University of Applied Sciences Augsburg. In research, teaching, and practice, he deals with issues related to data-driven \gls{bpm}, such as the extraction, quality assurance, and utilization of relevant data sources, the application of \gls{ai} methods for context-sensitive \gls{bpm}, as well as the management of \gls{ai} applications. The professorship is closely linked to the Branch Business \& Information Systems Engineering of the \gls{fit}, which contributes process mining expertise and a wide range of project experience from the \gls{cpi}.

Prof. Dr. Maximilian \textbf{R\"oglinger}\chairmarker{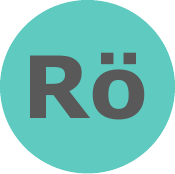} heads the \emph{Chair of Information Systems} at the University of Bayreuth. He works in research, teaching, and practice at the interface of customer, process, and IT, as well as in the field of digitalization. As part of the research network, he and his team focus on (automated) process improvement and process digitalization. The chair is closely linked to the Branch Business \& Information Systems Engineering of the \gls{fit}, where Prof. Dr. R\"oglinger holds a leading position. Together with his team, he contributes process mining expertise and a wide range of project experience from the \gls{cpi}.

Prof. Dr. Agnes \textbf{Koschmider}\chairmarker{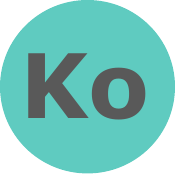} is professor of \emph{Information Systems and Process Analytics} at the University of Bayreuth. Further, she is spokeswoman of the DFG research unit FOR 5495 SOURCED "Process Mining on Distributed Event Sources" and co-applicant of NFDIxCS. She and her team research on methods for the data-driven analysis and explanation of processes as well as process behavior predictions based on \gls{ai} methods. Another research focus lies on methods for privacy-preserving analysis of process data as well as on the development of a data pipeline for efficient processing of different types of raw data, which allows to give novel insights to other disciplines. 

Research and teaching at the \emph{Chair of Databases and Information Systems} at the University of Bayreuth, headed by Prof. Dr. Stefan \textbf{Jablonski}\chairmarker{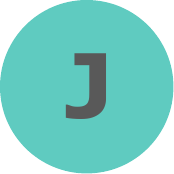}, focus on the interface of information systems and \gls{bpm}. A pivotal research aspect is the adept utilization of \gls{nlp} techniques in \gls{bpm}, exemplified by the automatic extraction of process models from natural language texts. A second research focus is the development of techniques for enhancing data-driven process execution and monitoring. This involves the development of methods for overcoming data scarcity as well as the creation of recommender systems tailored for process execution.

Prof. Dr. Martin \textbf{Matzner}\chairmarker{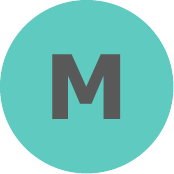} leads the \emph{Chair of Digital Industrial Service Systems} under the Institute of Information Systems at the Friedrich-Alexander-Uni\-vers\-it\"at Er\-lan\-gen-N\"urn\-berg. His research, projects, and teaching activities revolve around the four pillars of service systems, information management, decision support systems, and \gls{bpm}. Combining \gls{bpm} and decision support systems, the team investigates techniques and methods to identify and classify organizational routines, find potential improvements of business processes, and steer process executions into beneficial directions. Hence, the chair's mission is to improve decision-making and information processing in organizations for the mutual benefit of organizations, individuals, and society.

Prof. Dr. Stefanie \textbf{Rinderle-Ma}\chairmarker{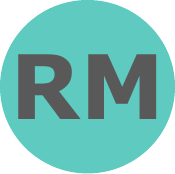} holds the \emph{Chair of Information Systems and Business Process Management} at the Technical University of Munich. Her research, teaching, and projects deal with process-oriented information systems, flexible and distributed process technologies, compliance management, as well as process and production intelligence. The main goal of her research is the development of new concepts and technologies to drive digitalization and automation through process science in various areas, including production, logistics, and the medical domain.

\end{document}

%% file: acronyms.tex
\newacronym{ai}{AI}{artificial intelligence}
\newacronym{bpm}{BPM}{business process management}
\newacronym{cpi}{CPI}{Fraunhofer Center for Process Intelligence}
\newacronym{etl}{ETL}{extract-transformation-load}
\newacronym{gdpr}{GDPR}{general data protection regulation}
\newacronym{genai}{GenAI}{generative artificial intelligence}
\newacronym{it}{IT}{information technology}
\newacronym{iot}{IoT}{Internet of Things}
\newacronym{iiot}{IIoT}{Industrial Internet of Things}
\newacronym{kpi}{KPI}{key performance indicator}
\newacronym{llm}{LLM}{large language model}
\newacronym{ml}{ML}{machine learning}
\newacronym{nlp}{NLP}{natural language processing}
\newacronym{pii}{PII}{process improvement and innovation}
\newacronym{piis}{PIIS}{process improvement and innovation systems}
\newacronym{rpa}{RPA}{robotic process automation}
\newacronym{crm}{CRM}{customer relationship management}
\newacronym{xai}{XAI}{explainable artificial intelligence}
\newacronym{erp}{ERP}{enterprise resource planing}
\newacronym{pais}{PAIS}{process-aware information systems}
\newacronym{dl}{DL}{deep learning}
\newacronym{fit}{Fraunhofer FIT}{Fraunhofer Institute for Applied Information Technology FIT}
\newacronym{dproma}{DProMa}{Research Network on Data-driven Business Process Management}